\documentstyle[12pt,epsf]{article}

\newcommand{\be}{\begin{equation}}
\newcommand{\ee}{\end{equation}}

\input epsf

\begin{document}

\begin{flushright}
ITEP-PH-11/99
\end{flushright}

\begin{center}
\large
{\bf The quark and gluon condensates and low-energy QCD theorems
in a magnetic field}
\end{center}

\vspace{5mm}

\begin{center}
{\bf N.O. Agasian
\footnote{e-mail: agasyan@heron.itep.ru}
}
\vspace{0.3cm}

and
\vspace{0.3cm}

{\bf I.A. Shushpanov
\footnote{e-mail: shushpan@heron.itep.ru}
}
\vspace{0.5cm}

\it{Institute for Theoretical and Experimental Physics,
B. Cheremushkinskaya 25, Moscow 117218, Russia}

\end{center}

\vspace{1.cm}
\centerline{November 1999}
\vspace{1cm}

\begin{abstract}
The low-energy QCD theorems are generalized in the presence of 
a constant magnetic field $H$. 
Two-loop approximation for the vacuum 
energy density in the framework of the chiral
perturbation theory was obtained and  the quark and gluon 
condensates were found as the functions of $H$.
\end{abstract}

\section{Introduction.}
The low-energy theorems, playing an important role in the understanding 
the vacuum state properties in quantum field theory,
were discovered almost at the same as quantum field methods appeared
in particle physics (see, for example, Low theorems $\cite{low}$). 
In QCD, these theorems were discovered in the beginning of eighties
$\cite{LT}$. These theorems, being derived from the very general 
symmetrical considerations 
and not depending on the details of confinement mechanism,
sometimes gives information which is not easy to obtain in another
way. Also, they can be used as "physically sensible" restrictions
in the constructing of effective theories. Recently, they were
generalized to finite temperature and chemical potential case
$\cite{LWT}$.

The behavior of the quark condensate in the presence of a magnetic
field was studied in Nambu-Jona-Lasinio model earlier $\cite{NJL}$.
For QCD, the analogous investigation in one-loop approximations
was done in $\cite{SS}$. It was found that 
the quark condensate grows with the increase of the magnetic field
$H$ in both cases. It implies that naive analogy with 
superconductivity, where the order parameter vanishes at some 
critical field, is not valid here.

In this paper, we generalize the low-energy theorems
to the case of the presence of the constant magnetic field.
When the field is weak (compared to the characteristic
hadronic scale), two-loop formulae for the vacuum energy density, 
the quark and gluon condensates as the functions of $H$ were 
obtained basing on the chiral perturbation theory (ChPT).
Gluons do not carry electric charge, nevertheless,
virtual quarks
produced by them interact with electromagnetic field
and lead to the changes in the gluon condensate.

\section{Low-energy theorems in the magnetic field.}

The Euclidean version of QCD partition function
in the magnetic 
field $A_\mu$ can be written as follows
\be
\label{statsum}
Z=\exp \left \{-\frac{1}{4} \int d^4 x F^2_{\mu\nu} \right \}
 \int [ {\cal D} B^a_\mu ]  [ {\cal D} \bar q_f ]  [ {\cal D} q_f ]
\exp \{ -\int d^4 x L_{QCD} \},
\ee
where QCD Lagrangian in the background field is 
\be
L_{QCD}= \frac{1}{4 g_0^2} ( G^a_{\mu \nu} )^2
+\sum_f  {\bar q_f} \left [ \gamma_\mu
( \partial_\mu - i Q_f e A_\mu -i \frac{\lambda^a}{2} B^a_\mu)
  +m_f \right ] q_f ,
\ee
Here $Q_f$  is the charge matrix of quarks in the flavor space
and we have suppressed gauge fixed and Faddeev-Popov terms
since they are unessential here. 
The vacuum energy density is defined in the usual way,
$V_4 \epsilon_v (H,m_f)=-\ln Z$.
In the constant magnetic field, the chiral condensate in 
the chiral limit $(m_f \rightarrow 0)$ 
is given by
\be
\label{qq}
\langle \bar q_f  q_f  \rangle (H)=\left. \frac{\partial
\epsilon_v (H,m_f) }{\partial m_f}
\right |_{m_f=0}
\ee
From ($\ref{statsum}$) follows that the gluon condensate
$\langle G^2 \rangle \equiv \langle (G^a_{\mu\nu})^2 \rangle $
can be written in the following form
\be
\label{condGG}
\langle G^2 \rangle =4 \left. \frac{\partial
\epsilon_v (H,m_f)}{\partial (1/g^2_0 )} \right |_{m_f=0}
\ee
The effect of dimensional transmutation produces a new dimensionful
nonperturbative parameter
\be
\label{Lambda}
\Lambda = M \exp \left \{ \int^\infty_{\alpha_s (M) }
\frac{d\alpha_s}{\beta (\alpha_s)}  \right \},
\ee
where $M$ - is ultraviolet cut-off, $\alpha_s=g_0^2 /4\pi$,
and $\beta(\alpha_s)=d\alpha_s (M) /d\ln M$ is Gell-Mann - Low
function.

The system, corresponding to partition function ($\ref{statsum}$), 
contains two dimensionful parameters $M$ and $H$ in the chiral limit 
$(m_f=0)$. As the vacuum energy density is an observable quantity,
it is also renorm-invariant one and its anomalous dimension is
zero. Therefore, $\epsilon_v$ has only normal (canonical) dimension
which equals four. Using the renorm-invariance of $\Lambda$,
we can write the most general expression for $\epsilon_v$
\be
\label{Erazm}
\epsilon_v=\Lambda^4 f(H/\Lambda^2),
\ee
where f is some function.
Then, it is easy to get from Eqs. ($\ref{Lambda}$) and 
($\ref{Erazm}$) that
\be
\label{GGG}
\frac{\partial \epsilon_v}{\partial (1/g_0^2)}=
\frac{8\pi \alpha_s^2}{\beta (\alpha_s)}
\left ( 2-H\frac{\partial }{\partial H} \right ) \epsilon_v,
\ee
and from Eq. ($\ref{condGG}$), we can find the relation between
$\epsilon_v$ and the gluon condensate 
\be
\label{GG}
\langle G^2 \rangle (H)=\frac{32\pi \alpha_s^2 }{\beta(\alpha_s)}
\left ( 2-H\frac{\partial}{\partial H} \right ) \epsilon_v
\ee
If we set $H=0$ we would reproduce the well-known expression for 
nonperturbative density of the vacuum energy in the chiral limit.
In one-loop approximation
($\beta=-b_0 \alpha_s^2/2\pi$, $b_0=(11 N_c-2 N_f)/3 $),
it reads 
\be
\epsilon_v=-\frac{b_0}{128\pi^2} \langle G^2 \rangle .
\ee

Let us now derive the low-energy theorems in the presence of
magnetic field. We can iterate the procedure described above
by taking $n$ further derivatives of Eq. ($\ref{condGG}$)
$$ (-1)^n \left ( 4-2H\frac{\partial}{\partial H}
\right )^{n+1} \epsilon_v =
\left ( 2H\frac{\partial}{\partial H} -4
\right )^n \langle \sigma \rangle =
$$
\be
=\int d^4 x_n \cdots d^4 x_1 \langle \sigma (x_n) \cdots
\sigma (x_1) \rangle_c.
\ee
Here 
\be
\sigma (x)= \theta_{\mu \mu} (x) =
\frac{\beta (\alpha_s)}{16 \pi \alpha^2_s} (G^a_{\mu\nu} (x) )^2,
\ee
and the subscript '$c$' means that only connected diagrams
are to be included. Proceeding in the same way, it is possible to obtain
the similar theorems for renorm-invariant operator $O(x)$
which is built from quark and/or gluon fields
\be
\left ( 2 H\frac{\partial}{\partial H}-d
\right )^n \langle O \rangle =
\int d^4 x_n \cdots d^4 x_1 \langle \sigma (x_n) \cdots
\sigma (x_1) O (0)\rangle_c ,
\ee
where $d$ - is canonical dimension of operator $O$. If operator
$O$ has also anomalous dimension, the appropriate 
$\gamma$-function should be included.

\section{Vacuum energy in the magnetic field.}

The obtained results permit us to calculate the condensates 
in the chiral limit as functions of $H$, once the vacuum energy 
density is known. To calculate $\epsilon_v$, we
have to consider the loops in the magnetic field. 
When the field is weak, 
$eH <<\mu^2_{hadr} \sim (4\pi F_\pi)^2$, characteristic momenta
in loops are small and theory is adequately described by
the low-energy effective Lagrangian $\cite{chpt}$ which
admits an expansion in powers of the external momenta
(derivatives) and the quark masses
\footnote{Note that the chiral limit means $eH>>M^2_\pi$.}
\be
\label{L}
L_{eff}=L^{(2)}+L^{(4)}+L^{(6)}+\cdots
\ee
The leading term $L^{(2)}$ in ($\ref{L}$)
is similar to the Lagrangian
of nonlinear $\sigma$-model in external field  $V_\mu$
\be
\label{CL2}
L^{(2)}=\frac{F^2_\pi}{4} {\rm Tr} (\nabla_\mu U^+ \nabla_\mu U)+
\Sigma {\rm Re Tr } ({\cal M} U^+ ),
\ee
$$
\nabla_\mu U = \partial_\mu U - i \left [ U, V_\mu \right ] .
$$
Here $U$ stands for an unitary $SU(2)$ matrix, $F_\pi=93$~Mev 
is the pion decay constant and parameter $\Sigma$ 
has the meaning of the quark condensate
$\Sigma=
|\langle \bar {u} u \rangle |=|\langle \bar {d} d \rangle |.$
The external Abelian magnetic field $H$, aligned along $z$-axis
corresponds to $V_\mu (x) = e (\tau^3 /2) A_\mu  (x)$,
where vector potential $A_\mu$ may be chosen as 
$A_1 (x) =-H x_2$.

The difference of light quark mass $m_u -m_d$ enters
in the effective Lagrangian ($\ref{L}$) only quadratically.
To calculate the fermion condensate, we need to take only first
derivative over the quark mass before going to the chiral limit.
In the case of the gluon condensate, we can put $m_u = m_d =0$ 
from
the very beginning. It means that we can take mass matrix
to be diagonal ${\cal M}=m {\hat I}$.

In the leading approximation the Lagrangian ($\ref{CL2}$) 
coincides with the Lagrangian of scalar electrodynamics.
In one-loop approximation, the result for the vacuum energy density
in this theory was found long time ago by Schwinger $\cite{shw}$
\be
\label{E1loop}
\epsilon_v^{(1)} (H) = -\frac{1}{16\pi^2}\int_0^\infty
\frac{ds}{s^3} e^{-M_\pi^2 s} \left [ \frac{eHs}{\sinh eHs} -1
\right ]
\ee
Using Eqs. ($\ref{qq}$), ($\ref{GG}$) and Gell-Mann - Oakes - Renner
relation (GMOR) $2m \Sigma =F^2_\pi M^2_\pi$, 
we can get one-loop formulae for the condensates in
the constant magnetic field
\be
\label{Sigma1}
\Sigma(H)=\Sigma  \left [ 1+\frac{eH \ln 2}{(4\pi F_\pi)^2}
\right ] ,
\ee
\be
\label{GG1}
\langle G^2 \rangle (H)=
\langle G^2 \rangle +\frac{\alpha_s^2}{3 \pi \beta (\alpha_s)}
(eH)^2.
\ee
The expression ($\ref{Sigma1}$) for the quark condensate was
obtained in $\cite{SS}$ earlier.
In the chiral perturbation theory in the magnetic field, 
the parameter of the expansion is $eH/(4\pi F_\pi)^2$. 
To find $\epsilon_v$ to the next order or the expansion,
we have to take into account two-loop diagrams with the vertices
from $L^{(2)}$, one-loop diagrams with one vertex from $L^{(4)}$ 
and a tree contribution arising from $L^{(6)}$. 
The corresponded Feynman graphs for $\epsilon^{(2)}_v$ 
are depicted in Fig. 1.

\begin{figure}
\centerline{\epsfbox{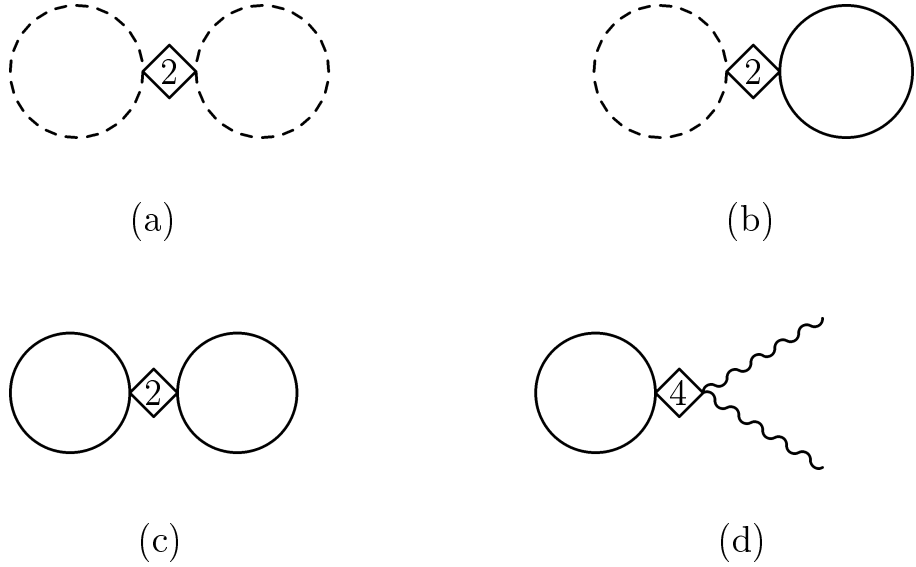}}
\caption{The loop diagrams contribute to vacuum energy density 
to the second order of ChPT. Solid lines correspond to the charged pions
and dashed lines correspond to the neutral pions.}
\end{figure}

To perform the calculations of these diagrams, we should
expand $L^{(2)}$ up to four-pion vertices.
The matrix $U$ can be parameterized in many ways. We choose
Weinberg parameterization
\be
\label{L2}
U=\sigma + \frac{i \pi^a \tau^a}{F_\pi},\quad
\sigma^2+\frac{\vec \pi^2}{F^2_\pi}=1.
\ee
Then the expansion of $L^{(2)}$ is
$$L^{(2)}=\frac{1}{2} (\partial_\mu \pi^0)^2 -\frac{M^2_\pi
(\pi^0)^2}{2}+ (\partial_\mu \pi^+ +ie A_\mu \pi^+) (\partial_\mu \pi^- -ie
A_\mu \pi^-) -M^2_\pi \pi^+ \pi^-
$$
\be +\frac{1}{2F^2_\pi} [\pi^0
\partial_\mu \pi^0 + \partial_\mu (\pi^+ \pi^- )]^2 -\frac{M^2_\pi}{8
F^2_\pi} [2\pi^+ \pi^- + (\pi^0)^2]^2 ,
\ee
where we have introduced the fields of the charged and the 
neutral pions
\be
\pi^0=\pi^3, \quad \pi^{\pm}
= \frac{1}{\sqrt 2} (\pi^1 \pm i\pi^2).
\ee
We need also the expression of a charged pion propagator in a magnetic
field. It can be inferred from the results of Ref. $\cite{prop}$
where an explicit expression for the fermion propagator 
in a magnetic field at nonzero chemical potential $\mu$
has been found. After some simplifications, we arrive at the
following formula for the Euclidean scalar propagator
\be
D^H (x,y)=\Phi(x,y) \int \frac{d^4
k}{(2\pi)^4} e^{ik(x-y)} D^H (k),
\ee
where $\Phi(x,y)=\exp\{ie \int^x_y A_\mu
(z) dz_\mu\}$ is the phase factor and the integration is done 
along the straight line connecting $x$ and $y$, and 
\be 
D^H (k)=
\int_0^\infty \frac{ds}{\cosh (eHs)}
\exp \left \{-s \left (
k^2_\parallel+k^2_\perp \frac{\tanh eHs}{eHs} + M^2_\pi \right )
\right \}
\ee
with $k^2_\parallel=k^2_3 + k^2_4$, $k^2_\perp=k^2_1 + k^2_2$.

The contribution of the graph in Fig. 1a to $\epsilon_v$ is
\be
\label{ea}
\epsilon^{(2)}_v [{\rm Fig. 1a}]=-\frac{M^2_\pi}{8 F^2_\pi} D^2 (0),
\ee
where $D(0)$ is a free propagator of the scalar particle at
coinciding points which can be written 
in dimensional regularization scheme as
\be
\label{D0}
D(0)=\int \frac{d^d k}{(2\pi)^d} \frac{1}{k^2+M^2}=
2 M^2 \left (
\lambda + \frac{1}{32 \pi^2} \ln \frac{M^2}{\mu^2} \right ) ,
\ee
where $\mu$ is the mass parameter of the regularization and
a singular term in ($\ref{D0}$)
is explicitely separated out as
\be
\lambda=\frac{\mu^{d-4}}{16 \pi^2} \left [ \frac{1}{d-4} -\frac{1}{2}
(\ln 4\pi -\gamma_E +1)\right ] .
\ee
The correction ($\ref{ea}$) 
can be absorbed to $\epsilon_v$ at $H=0$ and does not cause the shifts 
of the condensates by magnetic field.

The correction to $\epsilon_v$, 
coming from the next diagram Fig. 1b,
is 
\be
\label{eb}
\epsilon^{(2)}_v [{\rm Fig. 1b}]
=\frac{M^2_\pi}{2 F^2_\pi} D (0) D^H (0),
\ee
where $D^H (0)=D^H (x,x)$. According to GMOR, the result in
($\ref{eb}$) is proportional to the square of the quark mass and
does not change the condensates in the chiral limit.

Performing the calculation of the diagram Fig. 1c, we get
\be
\label{ec}
\epsilon^{(2)}_v [{\rm Fig. 1c}]=\frac{1}{F^2_\pi} D^H (0)
\int \frac{d^d k}{(2\pi)^d} (k^2+M_\pi^2) D^H (k).
\ee
This expression contains a quartic divergence. However,
by virtue of the dimensional regularization identity
$\int d^d k=0$, this divergence can be ignored
\footnote{For other regularizations, quartic divergence
is removed only if a nontrivial measure for the group 
integration is taken into account. The examples of explicit
calculations can be found in $\cite{measure}$.}.
Subtracting one from the integrand in
($\ref{ec}$) and going to a limit
$d \rightarrow 4$, we find that 
$\epsilon^{(2)}_v [Fig. 1c] = 0$.
So, we arrive at the conclusion that two-loop diagrams do not 
give the correction to $\epsilon_v$, linear in the quark mass
and, therefore, do not shift the condensates.

In this order of expansion in $eH/(4\pi F_\pi)^2$,
there are some diagrams with vertices from $L^{(4)}$ 
besides the diagrams discussed above.
As the momentum of a constant field is zero, only the two terms
from the general form  of $L^{(4)}$ are left
\be
L^{(4)}=-\frac{2 l_5}{F_\pi^2} (e F_{\mu\nu})^2 \pi^+ \pi^- -
\frac{2i l_6}{F_\pi^2} e F_{\mu\nu} [\partial_\mu \pi^-
\partial_\nu \pi^+
+ ie A_\mu \partial_\nu (\pi^+ \pi^- ) ],
\ee
where the phenomenological constants (infinite) $l_5$ ¨ $l_6$ 
were defined in $\cite{chpt}$.
The corresponding diagrams are drawn in Fig. 1d.
The calculation is straightforward and gives
\be
\label{E2loop2}
\epsilon_v^{(2)} [{\rm Fig. 1d}] = \frac{2(eH)^2}{F^2_\pi} 
(2 l_5 -l_6) D^H (0).
\ee
Although the constants $l_5$ ¨ $l_6$ are infinite, their combination, 
arising in ($\ref{E2loop2}$) is finite $\cite{chpt}$
\be
2l_5-l_6= \frac{1}{96\pi^2} (\bar l_6 - \bar l_5),
\ee
and $\bar l_6 - \bar l_5 \approx 2.7$.

Total expression $L^{(6)}$, given by chiral symmetry,
is rather complicated $\cite{LP6}$. However, for our purposes, 
the only one term is important which can be taken in the following
form, convenient for us
\be
\label{L6}
\label{d}
L^{(6)}=\frac{80d}{9 F_\pi^4} (e F_{\mu\nu})^2 \Sigma
{\rm Re Tr } \{ {\cal M} U^+ \}
\ee
Here $d=d^r (\mu) + const \cdot \lambda$ and $\lambda$
involves a pole $\mu^{(d-4)} /(d-4)$.
The divergent part in ($\ref{d}$) is canceled by the pole terms,
stemming from one-loop diagrams with vertices from $L^{(4)}$.
Numerical value of $d^r$ can be inferred from the results
obtained in $\cite{d6}$ where the process 
$\gamma\gamma \rightarrow \pi^0 \pi^0$ has been studied.
In the notations of
$\cite{d6}$, $d^r$ can be rewritten as
\be
d^r=\frac{9}{320} (\bar a_1 +2 \bar a_2 + 4 \bar b ),
\ee
where $\bar a_1$, $\bar a_2$ and $\bar b$ have the meaning of the 
coefficients at different tensor and mass structures in
the amplitude of the process $\gamma\gamma \rightarrow \pi^0 \pi^0$. 
The numerical values of $\bar a_1$, $\bar a_2$ and $\bar b$ 
have been obtained by vector, tensor and scalar resonances 
saturation of the amplitude.
$\cite{d6}$. It turns out that only the exchange by scalar mesons 
gives the contribution to $d^r$ and
\be
\label{numd}
d^r(\mu \sim 0.75 \quad {\rm Gev})\approx \pm 4 \cdot 10^{-6}
\ee
The scalar meson-photons and meson-pions coupling constants 
enter
in the experimentally observable decay widths quadratically,
since in the effective chiral Lagrangian they emerge linearly.
Hence, it easy to see that the sign of $d^r$ remains
undetermined.

Now we are in the position to get a finite result for 
$\epsilon_v$.
To find the quark condensate, one should keep only
first order terms in $M_\pi^2 /eH$ expansion
$$
D^H (0) = [D^H (0) - D (0) ] + D (0) \approx
-\frac{eH \ln 2}{16 \pi^2}
$$
\be
+\frac{M_\pi^2}{16 \pi^2} \left [ \ln\frac{eH}{M^2_\pi} +C \right ]
+2 M_\pi^2 \left [
\lambda + \frac{1}{32 \pi^2} \ln \frac{M_\pi^2}{\mu^2} \right ] ,
\ee
where $C$ - is slowly varying function of $eH/M_\pi^2$ and
$C(0)\approx -0.2$.
Adding up the obtained results, we arrive at the final answer for
$\epsilon_v$ in two-loop approximation
$$\epsilon_v (H)=\epsilon_v (0) + \epsilon^{(1)}_v (H)+
\frac{1}{48\pi^2}
\frac{(eH)^2}{(4 \pi F_\pi)^2} (\bar l_6 -\bar l_5)
\left \{ -eH \ln 2  +M_\pi^2
\left [ \ln\frac{eH}{\mu^2} +C \right ] \right \}  $$
\be
\label{Evac}
-\frac{160 d^r (\mu)}{9 F_\pi^2} (eH)^2 M_\pi^2.
\ee
Using ($\ref{qq}$), ($\ref{GG}$) and GMOR relation, it is easy
to find how the quark condensate depends upon $H$
$$
\Sigma(H)=\Sigma \left \{ 1+\frac{eH}{(4\pi F_\pi)^2} \ln 2 -
\frac{1}{3}
\frac{(eH)^2}{(4 \pi F_\pi)^4} \left [
(\bar l_6 -\bar l_5)
\left ( \ln\frac{eH}{\mu^2} +C \right ) \right . \right .
$$
\be
\label{qqf}
\left . \left . - \frac{160 (4\pi)^4 }{3 } d^r (\mu)
\right ] \right \}.
\ee
Further, we have for the gluon condensate 
\be
\langle G^2 \rangle (H)=\langle G^2 \rangle +
\frac{\alpha_s^2 }{3\pi \beta(\alpha_s)} (eH)^2
\left [ 1+2\frac{eH}{(4\pi F_\pi)^2}
(\bar l_6 -\bar l_5) \ln2 \right ].
\ee
It follows from asymptotic freedom that $\beta (\alpha_s) < 0$ 
and, consequently, the gluon condensate drops when the magnetic
field arises. In one loop approximation for QCD $\beta$-
function ($\beta (\alpha_s) = -b_0 \alpha_s^2 /2\pi$ ),
this decrease is equal to
\be
\Delta \langle G^2 \rangle = -\frac{2\pi}{3 b_0}
(eH)^2 \left [1+2 \frac{eH}{(4\pi F_\pi )^2 } (\bar l_6 - \bar l_5)
\ln 2 \right ]
\ee
Let us introduce the dimensionless variable $x=eH/(4\pi F_\pi )^2$.
Substituting  the found numerical values
into Eq. ($\ref{qqf}$),
we can write the quark condensate as a function of $x$
\be
\Sigma(x)/\Sigma=1+x\ln 2 -a x^2 \ln x -b x^2.
\ee
Here $a\simeq 0.9$ and $b\simeq 0.65 \pm 1.77$, 
where, for the coefficient $b$, we have used two values 
of $d^r$ from ($\ref{numd}$). The behavior of condensates
$\langle G^2 (x) \rangle / \langle G^2 \rangle $ and
$\Sigma (x)/\Sigma$ (for two different values of $b$) is shown
in Fig. 2. If $d^r > 0$ then the quark condensate starts to decrease
at the magnitude of the magnetic field $x > 0.23$. 

\begin{figure}
\centerline{\epsfbox{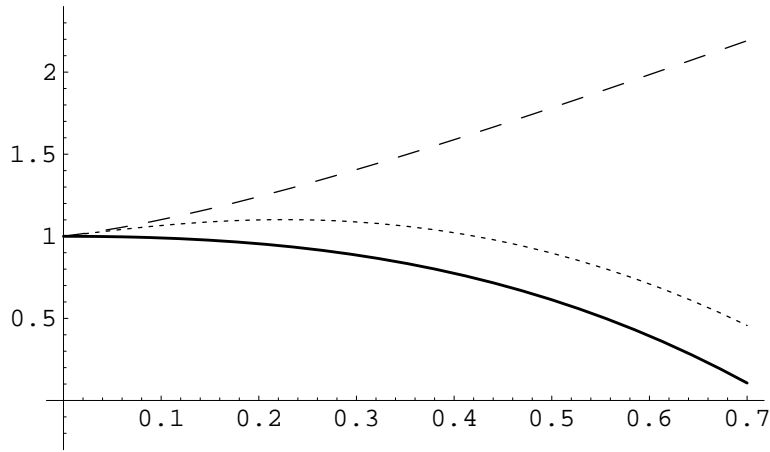}}
\caption{The dependence of the gluon condensate 
$\langle G^2 (x) \rangle / \langle G^2 \rangle $
(solid line) and
the quark condensate $\Sigma (x)/\Sigma$
(dotted line corresponds to $d^r > 0$ and dashed line corresponds
to $d^r<0$) upon $eH/(4\pi F_\pi)^2$.}
\end{figure}

\section{Conclusions}
We have generalized the low-energy QCD theorems
in the presence of the constant magnetic field $H$. 
Basing on ChPT, we have calculated the vacuum energy density
in two-loop approximation and found the dependence of 
the quark and gluon condensates upon the magnitude of $H$.
It is shown that the gluon condensate diminishes with the increasing 
of $H$ while
the behavior of the quark condensate
crucially depends on the sign of $d^r$. It decreases if we choose
positive sign and increases in the opposite case. Note
that possible decreasing of the condensate $\Sigma$
happens in the region of the applicability of ChPT
$eH/(4\pi F_\pi)^2 < 1 $. As it was already mentioned, it is not
possible to determine the sign of $d^r$ from experimental data.
We will not bring here various speculations concerning the behavior
of $\Sigma$ in the magnetic field. We would like only to say
that the question about the sign of $\Sigma$ changing in
two-loop level would be resolved if the phenomenological
constants of ChPT are derived from the first principles of QCD.

\vspace{5mm}

{\bf Acknowledgments}

We are grateful to B.L. Ioffe and Y.A. Simonov for useful
discussions and remarks.
This work was done under partial support of 
CRDF RP2-132 and RFFI 97-02-16131 grants.

\end{document}